\documentclass[8.5pt,twoside,twocolumn]{article}
\usepackage[super,sort&compress,comma]{natbib} 
\usepackage{mhchem}
\usepackage{times,mathptm}
\usepackage{sectsty}
\usepackage{balance} 
\usepackage{color}

\usepackage{graphicx} %eps figures can be used instead
\usepackage{lastpage}
\usepackage[format=plain,justification=raggedright,singlelinecheck=false,font=small,labelfont=bf,labelsep=space]{caption} 
\usepackage{fancyhdr}
\usepackage{epic}
\usepackage{color}
\begin{document}

\thispagestyle{plain}
\fancypagestyle{plain}{
%\fancyhead[L]{\includegraphics[height=8pt]{headers/LH.pdf}}
%\fancyhead[C]{\hspace{-1cm}\includegraphics[height=20pt]{headers/CH.pdf}}
%\fancyhead[R]{\includegraphics[height=10pt]{headers/RH.pdf}}
\renewcommand{\headrulewidth}{1pt}}
\renewcommand{\thefootnote}{\fnsymbol{footnote}}
\renewcommand\footnoterule{\vspace*{1pt}% 
\hrule width 3.4in height 0.4pt \vspace*{5pt}} 

\makeatletter 
\renewcommand\@biblabel[1]{#1}            
\renewcommand\@makefntext[1]% 
{\noindent\makebox[0pt][r]{\@thefnmark\,}#1}
\makeatother 
\renewcommand{\figurename}{\small{Fig.}~}
\sectionfont{\large}
\subsectionfont{\normalsize} 

\fancyfoot{}
%\fancyfoot[LO,RE]{\vspace{-6pt}\includegraphics[height=8.5pt]{LF.pdf}}
%\fancyfoot[CO]{\vspace{-6.5pt}\hspace{11.4cm}\includegraphics{RF.pdf}}
%\fancyfoot[CE]{\vspace{-6.6pt}\hspace{-12.7cm}\includegraphics{RF.pdf}}
%\fancyfoot[RO]{\footnotesize{\sffamily{1--\pageref{LastPage} ~\textbar  \hspace{2pt}\thepage}}}
%\fancyfoot[LE]{\footnotesize{\sffamily{\thepage~\textbar\hspace{4.4cm} 1--\pageref{LastPage}}}}
\fancyhead{}
\renewcommand{\headrulewidth}{1pt} 
\renewcommand{\footrulewidth}{1pt}
\setlength{\arrayrulewidth}{1pt}
\setlength{\columnsep}{6.5mm}
\setlength\bibsep{1pt}

\twocolumn[
  \begin{@twocolumnfalse}
\noindent\LARGE{\textbf{Flexoelectric switching in cholesteric blue phases}}
\vspace{0.6cm}

\noindent\large{\textbf{A. Tiribocchi$^{\ast}${$^{a}$}, M. E. Cates$^{a}$, G. Gonnella$^{b}$, D. Marenduzzo$^{a}$ and E. Orlandini\textit{$^{c}$}}}\vspace{0.5cm}
%Please note that \ast indicates the corresponding author(s) but no footnote text is required. 

\noindent\textit{\small{\textbf{Received Xth XXXXXXXXXX 20XX, Accepted Xth XXXXXXXXX 20XX\newline
First published on the web Xth XXXXXXXXXX 20XX}}}

\noindent \textbf{\small{DOI: 10.1039/b000000x}}
\vspace{0.6cm}
%Please do not change this text.

\noindent \normalsize{
We present computer simulations of the response of a 
flexoelectric blue phase network, 
either in bulk or under confinement, to an applied field. 
We find a transition in the bulk between the blue phase I disclination network 
and a parallel array of disclinations along the direction of the applied field.
Upon switching off the field, the system is unable to reconstruct the original 
blue phase but gets stuck in a metastable phase. Blue phase II is comparatively
much less affected by the field.
In confined samples, the anchoring at the walls and the geometry of the device
lead to the stabilisation of further structures, including field-aligned
disclination loops, splayed nematic patterns, and yet more metastable states.
Our results are relevant to the understanding of the switching dynamics for
a class of new, ``superstable'', blue phases which are composed of bimesogenic
liquid crystals, as these materials combine anomalously large flexoelectric 
coefficients, and low or near-zero dielectric anisotropy. 
}
\vspace{0.5cm}
 \end{@twocolumnfalse}
  ]

\section{Introduction}

\footnotetext{\textit{$^{a}$~SUPA, School of Physics and Astronomy, University of Edinburgh, Mayfield Road, Edinburgh EH9 3JZ, UK}}
\footnotetext{\textit{$^{b}$~Dipartimento di Fisica and Sezione INFN di Bari, Universita' di Bari, I-70126 Bari, Italy}}
\footnotetext{\textit{$^{c}$~Dipartimento di Fisica e Astronomia and Sezione INFN di Padova, Universita' di Padova, Via Marzolo 8, 35131 Padova, Italy}}

The blue phases (BPs) of cholesteric liquid crystals provide a remarkable 
example of soft materials with a rare combination of highly nontrivial 
geometric and topological nature, and promising technological 
potential. 
BPs are three-dimensional networks of disclination lines, which appear due to 
the spontaneous tendency of a cholesteric liquid crystal to twist along more 
than one direction. The associated double twist regions lead to topological
frustration, as they cannot be used to smoothly tile the three-dimensional
space, so that defects, or disclination lines, are unavoidable. 
In two of the experimentally observed blue phases, BP I and BP II,
to which we restrict the present study, such disclinations form ordered 
lattices, with symmetries characterised by point groups normally with associated 
crystals (BP III is instead likely amorphous). The spatial period of 
the disclination lattices is in the submicrometer range, which endows them
with interesting optical properties -- for instance, a swelling or shrinking
of the unit cell would normally be linked to a change in the colour
in the sample.

A leap forward in blue phase based technology has been
made possible by the stabilisation of these disclination networks
over a range of over 60 K, including room temperature~\cite{psbp,hisakado_2004,Coles_Pivnenko:2005:Nature,colesnatmat:2012,Karatairi_et_al:2010:Phys_Rev_E}. 
This was a dramatic advance with respect to the previous framework, in
which BPs were typically only stable over a fraction of a degree.
As a consequence, it has now become possible to think of blue phases
as a possible basis for device or display technology, and this
has culminated in the recent fabrication of a blue-phase display device with 
very fast switching and response times~\cite{samsung}. Such switching
times are a natural consequence of the small scale of the unit cells which
need to rearrange when a field is applied. Another potential 
technological advantage of blue phases, which has not been exploited 
much in practice so far, is that they are associated with a complex
and rich free energy landscape, with several competing equilibria which
are ideal to build bistable or multistable devices. Such devices can retain
memory of two or more states even after the field used to reach these
has been switched off (this is a highly sought after property in devices
as it leads to huge savings in energy consumption)~\cite{dennis}.

While the stable BPs of Ref.~\cite{psbp} involved dispersing
polymers inside BPs, those of Ref.~\cite{Coles_Pivnenko:2005:Nature}, 
which provide the main motivation for our current work, were a one-component 
fluid of ``bimesogenic''~\footnote{To form a bimesogenic liquid crystal, 
one needs a compound which is made up by two polar mesogenic units
linked by a flexible spacer~\cite{colesflexo}.} 
liquid crystals, which are associated with very large
flexoelectric coefficients~\cite{colesflexo}.
Flexoelectricity in liquid crystals is a 
linear coupling between applied electric field and induced distortion.
This arises, for instance, because the microscopic shape of the molecules
(which can be e.g. pear-shaped) allows a coupling between splay (or bend)
and polarisation. The treatment of Ref.~\cite{castles}
shows that in flexoelectric materials this coupling leads
to a renormalisation of the elastic constants, which
effectively decrease and facilitates the formation of disclination lines
hence of BPs. Conversely, in materials with large
flexoelectric coefficients, elastic distortions readily can be created 
by an external electric field -- the flexoelectric coupling is a linear
coupling between electric field and order parameter variations, which
is allowed in the theory due to the nonsymmetric nature of the molecule. 
For low enough voltage, this contribution will dominate over the
familiar, quadratic, dielectric coupling between order parameter and 
electric field.

Because any-device oriented application of blue phases is likely to
use the stabilised versions of these materials, it is very important to 
understand the switching behaviour of BPs in response to a 
flexoelectric, rather than dieletric, coupling to an electric field, 
as this is likely to be the
dominant contribution at least when preparing BP samples as in 
Ref.~\cite{Coles_Pivnenko:2005:Nature}. Therefore, in this work our
goal is to elucidate the effect of flexoelectricity in the switching
of BP domains, or confined BPs, which can now be potentially used to build real 
devices. To this end, we use large scale three-dimensional lattice Boltzmann
simulations, which have already allowed in the past a careful study of the 
thermodynamic and flow properties of BPs~\cite{oliver,softmatter,domaingrowth,atsoft,prl}. Importantly, our approach correctly
incorporates hydrodynamic interactions which are known to affect the switching
pathway of liquid crystal devices. Our work provides a direct systematic
numerical investigation of the switching dynamics of cholesteric blue phases 
(both in bulk and under confinement) in response to a 
flexoelectric coupling with the applied electric field (we refer to
this phenomenon as ``flexoelectric switching'').

This paper is structured as follows. In Section II, we review the free energy
density on which the continuum model we use is built, taking care to derive
and discuss the extra terms which are due to flexoelectricity and which
will play a key role in our work. In Section III we present our results,
first focussing on the effect of bulk flexoelectricity on the unit cells
of a large BP sample, and then analysing the case of confined geometries.
Finally, we draw our conclusion in Section
IV, where we also compare the flexoelectric switching studied here with
previous works studying response of BPs to a dielectric coupling.

\section{Models and methods}

\subsection{Free energy and equations of motion}

In order to investigate the switching properties of flexoelectric blue phases, we consider the formulation of  
liquid-crystal hydrodyamics given by Beris and Edwards~\cite{beris}, generalized for cholesteric liquid crystals.  
In this formulation the equations of motion are written in terms of a tensorial order parameter, 
${\bf Q}$. The latter is related to the local direction of the individual molecules, ${\bf \hat{n}}$, by 
$Q_{\alpha\beta}=\langle \hat{n}_{\alpha}\hat{n}_{\beta}-\frac{1}{3}\delta_{\alpha\beta}\rangle$, where
the angular bracket denotes a coarse-grained average and the Greek indices label the Cartesian components of ${\bf Q}$.
The tensor ${\bf Q}$ is traceless and symmetric  and its largest eigenvalue, $2/3q$,  $(0 < q < 1)$, measures the magnitude of local order.  The equilibrium properties of the system are described by a Landau-de Gennes free energy density~\cite{degennes,chandrasekar,mermin}.
This comprises a bulk term (summation over repeated indices is implied hereafter),
\begin{equation}
f_b=\frac{A_0}{2} (1 - \frac {\gamma} {3}) Q_{\alpha \beta}^2  -  
          \frac {A_0 \gamma}{3} Q_{\alpha \beta}Q_{\beta
          \gamma}Q_{\gamma \alpha} + \frac {A_0 \gamma}{4} (Q_{\alpha \beta}^2)^2,
\end{equation}        
which describes a first-order transition from the isotropic to the chiral (cholesteric) phase, together with an elastic contribution which for cholesterics is~\cite{degennes}
\begin{equation}
f_d = \frac{K}{2} (\partial_\beta Q_{\alpha \beta})^2
+ \frac{K}{2} (\epsilon_{\alpha \zeta\delta }
\partial_{\zeta}Q_{\delta\beta} + \frac{4\pi}{p_0}Q_{\alpha \beta})^2.
\end{equation}
In the formula above $K$ is the elastic constant (here we are assuming the one-elastic constant approximation) and $p_0$ is the intrinsic helix pitch of the cholesteric. The tensor $\epsilon_{\alpha \zeta\delta}$ is the Levi-Civita antisymmetric third-rank tensor, $A_0$ is a constant (with units of pressure) and  $\gamma$  controls the magnitude of order  (it plays the role of an effective temperature or concentration according to whether the cholesteric liquid crystal is thermotropic or lyotropic). 

To focus on the effect of flexoelectric coupling
with an external field $E_{\alpha}$, we assume
that the liquid crystal has zero dielectric anisotropy and the
electric field is uniform throughout the sample. Flexoelectric
contributions can be divided into a bulk and a surface one. 
To account for the former, we add the following term to the
free energy density~\cite{alex3}:
\begin{equation}
f_E = \epsilon_{fl}^bQ_{\alpha\beta}(E_{\alpha}\partial_{\gamma}-E_{\gamma}\partial_{\alpha})Q_{\beta\gamma},
\label{e_electric}
\end{equation}
where   $\epsilon_{fl}^b>0$ is the  bulk flexoelectric constant.  
Note that each term of the flexoelectric coupling is linear in $E_{\alpha}$ and quadratic in ${\bf Q}$.
This is different from the standard dielectric coupling  where  the field couples linearly to the orientational order parameter
and quadratically with the electric field.
The reduced potential is $V_z=E_zL_z$, where $E_z$ is the component of the electric field along the $z$-direction
and $L_z$ the size of the sample in the same direction.

To describe the anchoring of the director field and the  flexoelectric interactions at the boundary surfaces,  the following surface contributions are added to the free energy density~\cite{degennes}
\begin{eqnarray}
f_s&=&\frac{W_0}{2} (Q_{\alpha\gamma} - Q_{\alpha\gamma}^0)^2+\frac{K}{2} \left[(\partial_{\alpha}Q_{\beta\gamma})^2 + (\partial_{\alpha}Q_{\alpha\gamma} )(\partial_{\beta}Q_{\beta\gamma})\right]\nonumber\\
&&+\epsilon_{fl}^s \left ( \partial_{\beta} Q_{\alpha\beta}\right ) E_{\alpha}.
\label{surf_freeen}
\end{eqnarray}
The first term is a quadratic contribution ensuring a soft pinning of the director field on the boundary surface,  along a direction $\hat{n_0}$. The parameter $W_0$ controls the  strength of the anchoring and 
$Q_{\alpha\gamma}^0=S_0 (n_{\alpha}^0n_{\gamma}^0-\delta_{\alpha\gamma}/3)$, where $S_0$ determines the magnitude of surface order. In the strong anchoring regime $W_0$ is practically infinite so that the director field is effectively fixed at the boundaries. In the weak anchoring regime a term proportional to $K$, describing the elastic distorsion at the surfaces,  is also important (see second term in  Eq.~(\ref{surf_freeen})) . Finally, the flexoelectric contribution at the boundary surfaces
is  represented by the last term of  Eq.~(\ref{surf_freeen})   where  $\epsilon_{fl}^s>0$ is  the surface flexoelectric constant. 
This  contribution depends linearly  on the gradients of ${\bf Q}$,  and becomes irrelevant in the strong anchoring regime (because the integral of the associated free energy density can be rewritten as a surface integral over the boundary). 

It is often convenient to express the free energy in terms of dimensionless quantities. This decreases
the number of parameters necessary to describe the system, providing a minimal set on which phase behaviour
and electric field effects depend. The parameters are
\begin{eqnarray}
\kappa &=& \sqrt{\frac{108 K q_0^2}{A_0 \gamma}}\\
\tau &=& \frac{27(1-\gamma/3)}{\gamma}+ \kappa^2\\
{\cal E}_{fl}^b&=&\sqrt{\frac{E_{\alpha}E_{\alpha}(\epsilon_{fl}^b)^2}{A_0K}}\\
{\cal E}_{fl}^s&=&\frac{\epsilon_{fl}^s|E|}{W}.
\end{eqnarray}

The reduced temperature $\tau$ multiplies the quadratic term of the 
dimensionless bulk free energy, whereas the gradient free energy term
is proportional to the chirality $\kappa$. This parameter can be 
considered as a measure of the twist present in the system; it is indeed defined
as a ratio between the bulk and the gradient free energies.
Lastly, parameters ${\cal E}_{fl}^b$ and ${\cal E}_{fl}^s$~\cite{alex3} 
measure the strength of bulk and surface flexoelectric couplings. Also notice 
that in strong anchoring regime $W$ is set to infinity 
and ${\cal E}_{fl}^s\simeq 0$.

The equation of motion for ${\bf Q}$  is~\cite{beris,perm3}
\begin{equation}
(\partial_t+{\vec u}\cdot{\bf \nabla}){\bf Q}-{\bf S}({\bf W},{\bf
  Q})= \Gamma {\bf H}
\label{Qevolution}
\end{equation}
where $\Gamma$ is a collective rotational diffusion constant. The first term of the left hand side of Eq.~(\ref{Qevolution}) 
is the material derivative describing the usual time dependence from advection by a fluid with velocity $\vec{u}$. For the tensor field ${\bf Q}$,
this is generalized by the term~\cite{beris}
\begin{eqnarray}\label{S_definition}
{\bf S}({\bf W},{\bf Q})
& = &(\xi{\bf D}+{\bf \omega})({\bf Q}+{\bf I}/3)+ ({\bf Q}+
{\bf I}/3)(\xi{\bf D}-{\bf \omega}) \nonumber \\
& - & 2\xi({\bf Q}+{\bf I}/3){\mbox{Tr}}({\bf Q}{\bf W})
\end{eqnarray}
where Tr denotes the tensorial trace, while
${\bf D}=({\bf W}+{\bf W}^T)/2$ and ${\bf \omega}=({\bf W}-{\bf W}^T)/2$
are the symmetric and the anti-symmetric part of the
velocity gradient tensor $W_{\alpha\beta}=\partial_\beta u_\alpha$. 
The presence of   ${\bf S}({\bf W},{\bf Q})$  is due to the
fact that the order parameter distribution can be both rotated and stretched by the fluid.  
This is a consequence of the rod-like geometry of the LC molecules. 
The  constant $\xi$ depends on the aspect ratio of the molecules forming a
given liquid crystal and controls whether the director field is flow-aligning in shear
flow ($\xi \geq 0.6$ for $\gamma=3$), creating a stable response, or flow-tumbling, which
gives an unsteady response ($ \xi \leq 0.6$ for $\gamma=3$).
The term on the right-hand side of  Eq.~(\ref{Qevolution})  describes the relaxation 
of the order parameter towards the minimum of the free energy. 
The molecular field ${\bf H}$ which  provides  the driving motion is given by 
\begin{equation}
{\bf H}= -{\delta {\cal F} \over \delta {\bf Q}}+({\bf
I}/3) {\mbox{Tr}}{\delta {\cal F} \over \delta {\bf Q}}.
\label{molecularfield}
\end{equation}
with ${\bf I}$ being the unit matrix and ${\cal F}$ the free energy.

The three-dimensional fluid velocity, $\vec u$, obeys the continuity equation,
\begin{equation}\label{continuity}
\partial_t \rho + \partial_{\alpha} ( \rho u_{\alpha}) = 0
\end{equation}
and the Navier-Stokes equation, 
\begin{eqnarray}\label{navierstokes}
\rho(\partial_t+ u_\beta \partial_\beta)
u_\alpha = \partial_\beta (\Pi_{\alpha\beta})+
\eta \partial_\beta(\partial_\alpha
u_\beta + \partial_\beta u_\alpha),
\end{eqnarray}
where $\rho$ is the fluid density and $\eta$ 
is an isotropic viscosity. Note that the form of this equation is similar 
to that for a simple fluid. However, the details of the stress tensor,  $\Pi_{\alpha\beta}$, 
reflects the additional complication of  cholesteric liquid crystals hydrodynamics. 
This tensor is explicitly given by:
\begin{eqnarray}
\Pi_{\alpha\beta}= &-&P_0 \delta_{\alpha \beta} +2\xi
(Q_{\alpha\beta}+{1\over 3}\delta_{\alpha\beta})Q_{\gamma\epsilon}
H_{\gamma\epsilon}\\\nonumber
&-&\xi H_{\alpha\gamma}(Q_{\gamma\beta}+{1\over
  3}\delta_{\gamma\beta})-\xi (Q_{\alpha\gamma}+{1\over
  3}\delta_{\alpha\gamma})H_{\gamma\beta}\\ \nonumber
&-&\partial_\alpha Q_{\gamma\nu} {\delta
{\cal F}\over \delta\partial_\beta Q_{\gamma\nu}}
+Q_{\alpha \gamma} H_{\gamma \beta} -H_{\alpha
 \gamma}Q_{\gamma \beta}.
\label{BEstress}
\end{eqnarray}
Here $P_0$ is an isotropic background pressure.

\subsection{Numerical set up  and mapping to physical units}\label{Section2.1}

In this section we briefly describe the numerical aspects of our work.

The size of the simulation box is equal to $L_x=32, L_y= 32, L_z= 32$ lattice sites.
Note that the size of the unit cell which minimises the free energy is in general
different from (and typically larger than) the cholesteric pitch. 
Because we can only simulate an integer number of periodic unit cells
within a periodic simulation domain, in order to account for this extra degrees of freedom
we employ the following procedure. Keeping fixed
the size of the periodic domain, we rescale the 
size of the simulation box by a factor called \lq\lq redshift\rq\rq\footnote{This name refers to the shift, towards longer wavelengths, of the 
BPI Bragg back reflection with respect to that of the cholesteric phase~\cite{grebel}. In other words, the unit cell size of the blue phase is typically larger than 
the cholesteric half pitch.}~\cite{alex,alex2,gareth2}. 
In simulations with periodic boundaries in all directions, we have followed the recipe proposed in Ref.~\cite{alex3},
in which the redshift is calculated at every time step by minimizing the free energy. 
Since the free energy functional comprises terms up to quadratic order in gradients
of ${\bf Q}$, a  rescaling of the unit cell dimension, $K\rightarrow K/r$ (with $r$ the redshift),
modifies the gradient terms, that are in turn rescaled by a factor of $r$ per derivative. The free
energy can be rewritten as $f({\bf Q})=r^2a(\partial{\bf Q}^2)+rb(\partial{\bf Q})+c$, and the rescaling
factor $r^*$ that minimizes the free energy is $r^*=-b/2a$. In particular we assume a single 
redshift governing all the three dimensions~\cite{oliver}.  We found that the presence of
redshift does not affect the
switching dynamics of BPs cells~\cite{atsoft}. In simulations with homeotropic
or homogenous anchoring,  the redshift was set to one to avoid any change in the size of the confining cell.
According to the experience gained in the bulk simulations, allowing for a 
finite redshift 
can lead to quantitative, but not qualitative, differences in the dynamics.

The equilibrium structures of BPI and BPII in the absence of any field
were obtained by relaxing the free energy to its minimum value by numerically solving Eq.~(\ref{Qevolution}).
Initial configurations were set according to the approximate solutions given in Ref.~\cite{mermin},
where it was shown  that the topological character of the equilibrium defect structure at high chirality (where
the bulk and the external field free energy terms become negligible compared to 
the gradient free energy and an analytical solution is possible) 
is retained by reducing the chirality. These solutions are taken as initial conditions 
for the dynamic equations~\cite{alex}  (\ref{Qevolution}), (\ref{continuity}) and (\ref{navierstokes})
that are numerically  integrated by using an hybrid scheme in which  standard lattice Boltzmann
methods~\cite{softmatter} are used to update the Navier-Stokes equation while  Eq.~(\ref{Qevolution}) is 
solved by using  finite difference scheme based on a  predictor corrector algorithm. 
This approach has been already successfully tested for other systems such as
binary fluids~\cite{gonnella,Tiribocchi,gonnella2}, active nematics~\cite{Marenduzzo_et_al:2007:Phys_Rev_E,softmatter}
or blue phases~\cite{atsoft}. 
A typical simulation reported in this work was run in parallel on 8 nodes with MPI architecture, and
required about 5 days to be completed. 

In our calculations we fixed $\Gamma=0.3$, $\xi=0.7$ and 
$\eta=1.333$ as in previous numerical works~\cite{alex,atsoft}. 
We have also set $A_0=0.00308$, $K=0.005$, $\gamma=3.775$ for BPI, 
and $A_0=0.00075$, $K=0.005$, $\gamma=3.086$ for BPII. This choice guarantees we are
in the correct region of the phase diagram~\cite{alex}.
In order to get from simulations to physical units we follow the approach of Refs.~\cite{alex4,domaingrowth}.
Our parameters correspond to a blue phase with length scale fixed by the cholesteric pitch $p_0$, typically in the range of $100-500$nm~\cite{mermin}, and
(Frank) elastic constants equal to $\sim 10$ pN and $\sim 16$ pN for BPI and BPII respectively, as for a typical liquid crystalline material. 
One space and time LB unit correspond to 0.0125 $\mu$m and 0.0013 $\mu$s for BPI, and to 0.0125 $\mu$m and 0.0017 $\mu$s for BPII,
with a rotational viscosity equal to $1$ Poise.
The mapping of flexoelectric couplings can
be done by matching the previously defined dimensionless quantities, 
${\cal E}_{fl}^b$ and ${\cal E}_{fl}^s$, 
when defined in simulation and in physical units.
For instance, for BPII a value of ${\cal E}_{fl}^b=0.3$ (which is
in the middle of the range used in our runs) corresponds to a system
with (Frank) elastic constants equal to $\sim 10$ pN as before, with
a value of $\epsilon^{b}_{fl}\sim 10$ pC/m, subject to a field of $\sim 15$
 V/$\mu$m. These values are typical of liquid crystal devices; to
directly apply our results to this situation, we further need
to select a liquid crystal for which dielectric effects are negligible,
which is challenging at these field strengths -- a possibility is to 
use bimesogenic materials which have near-zero dielectric anisotropy. 
On the other hand, a value of the dimensionless surface flexoelectric coupling
of  ${\cal E}_{fl}^s=0.15$ may correspond to the same BPII material, 
when the anchoring at the surface is $W \sim 0.0007$ N/m. For BPI, a similar mapping applies.

Finally, to locate the defects in the simulation domain, we  used the  approach described in~\cite{atsoft}, 
where a lattice point is considered to belong to a defect  
if the local order at that point assumes values below a given threshold.
Here the  threshold is fixed to be  $60\%$ of the largest eigenvalue of ${\bf Q}$ in the bulk (i.e. away from all defects).

\section{Results}

\subsection{Flexoelectric effects in blue phase samples with periodic boundary conditions}

We now discuss the results obtained by looking at the switching dynamics of BPI and BPII unit cells 
with periodic boundary conditions. We will focus  on the deformations of the defect structure  observed
when a field is applied to the sample along  the $z$-direction ($[0,0,1]$ direction), only considering the effect of bulk flexoelectric couplings. The surface
flexoelectric contribution vanishes with these boundary conditions, and the dielectric contribution is
disregarded to focus on flexolectricity.   
For comparison, a single case in which the field is along the diagonal of the unit cell ($[1,1,1]$ direction) will be also considered,
to get a feel for the anisotropy of the response.

\begin{figure*}
\centerline{\includegraphics[width=18.cm]{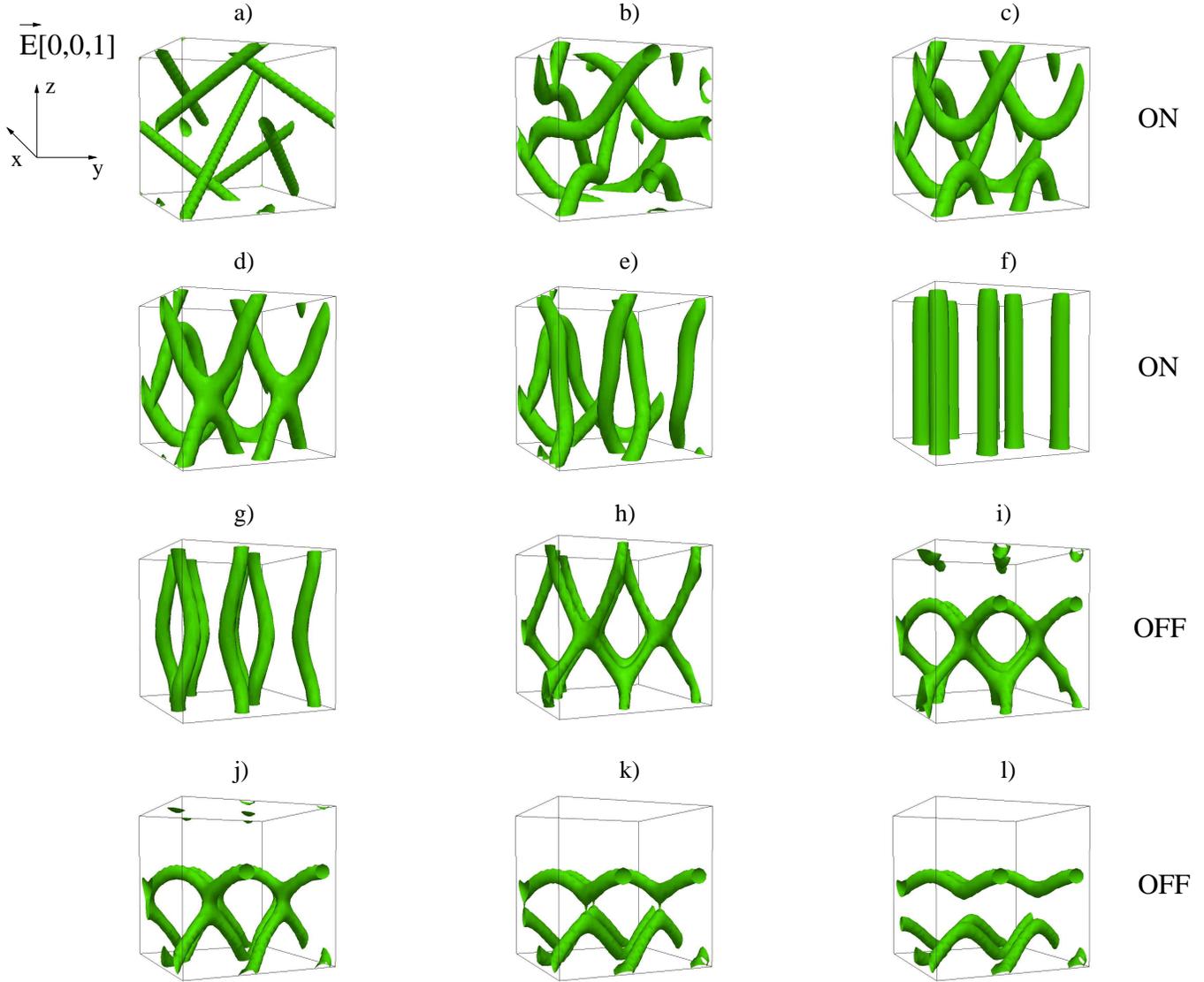}}
\caption{Defect dynamics of the BPI phase during the switching on ($a)-f)$)
-off ($g)-l)$)  in a simulation box of size $32\times 32\times 32$. Periodic boundary conditions are set in all directions.
The electric field, applied along the $z$-direction ($[0,0,1]$), is ${\cal E}_{fl}^b \simeq 0.5$. It is
switched off at $t=100\times 10^4$. One cycle is not enough to have a switchable device.
A further application of the field reveals the device is switchable between the states $f)$ and $l)$.}
\label{fig1}
\end{figure*}

BPs are arranged in a $L_x=L_y=L_z=32$ periodic unit cell. The initial equilibrium defect structures (see Fig.\ref{fig1}$a$ for BPI and Fig.\ref{fig2}$a$ for BPII) are obtained by minimizing  the free energy  by using exclusively the evolution equation   Eq.~(\ref{Qevolution}) ~\cite{alex,atsoft}. 
Notice that in this case the minimization  is driven exclusively  by the molecular field ${\bf H}$. Unless otherwise stated, hydrodynamic
interactions are switched on after this equilibration, and remain in place throughout the dynamical switching simulations. 

Fig.\ref{fig1} shows  the response of a periodic BPI unit cell under a bulk electric field applied along the $z$-direction.
The field strength chosen  (${\cal E}_{fl}^b \simeq 0.5$) is large enough to modify the initial defect structure in an appreciable way. 
When the field is switched on, disclinations start to bend and twist, although the typical BPI defect structure is still 
recognizable (Fig.\ref{fig1}$b$).  As time goes on these defect distortions become more and more pronounced resulting into  
strongly bent arcs  that span the entire unit cell and are aligned along the direction of the  field (Fig.\ref{fig1}$c$). 
Later arcs reorganize to form \lq\lq X-like\rq\rq defect junctions (Fig.\ref{fig1}$d$) that successively unbind 
at their center (Fig.\ref{fig1}$e$) giving rise to columnar-like disclinations parallel to the $z$-direction (Fig.\ref{fig1}$f$). 
Fig.\ref{fig1}  shows that  the flexoelectric switching dynamics under a strong field is extremely different from previously studied cases  in which only dielectric  effects have been considered~\cite{fukuda2,atsoft}. In the latter case
the disclinations twist (for small field) or reconnect to lead to an amorphous state, in stark contrast with the regular 
structures of parallel defect lines obtained here. For lower fields (provided these are still high enough to break the defect structure), the field-induced states 
are characterized by the same defect network, with the typical hexagonal pattern for the director profile. Interestingly, for those cases, 
during the switching on dynamics, double helical disclinations are observed 
initially, perpendicularly to the direction of the applied field -- these
later on merge and reorganize into the column-like pattern.
Similar textures have already been described in Ref.\cite{gareth2} as 
intermediate field-induced states, while in our simulations these are only
transient defect structures. These differences might be due to different 
values of bulk and elastic constants considered (for example, 
the theoretical description in~\cite{gareth2} relaxes the one
elastic constant approximation).

A crucial  point in the physics of  devices is the understanding of the switching off dynamics, i.e. the time evolution of the system when the field is switched off.  In particular one would like to answer  the following question:
Is the device switchable (i.e. does it recover the initial defect structure when the field is removed), or does it get stuck into a metastable state characterized by a different defect network ? 
In our experiment, as the flexoelectric coupling is switched off,  the disclination lines start to bend, approach one another (Fig.\ref{fig1}$g$), and merge into a connected structure that is  similar in shape to  the  \lq\lq X-like\rq\rq state observed during the switching on dynamics but  with thinner  disclinations (Fig.\ref{fig1}$h$). Later on defects reorganize themselves  into arcs that  first span the entire system  (Fig.\ref{fig1}$i$), then gradually stretch (Fig.\ref{fig1}$j$)  and finally split  forming long undulated defect lines (Fig.\ref{fig1}$k$ and $l$), some of them in proximity of the center of unit cell, others closer to the boundaries.

\begin{figure*}
\centerline{\includegraphics[width=18.cm]{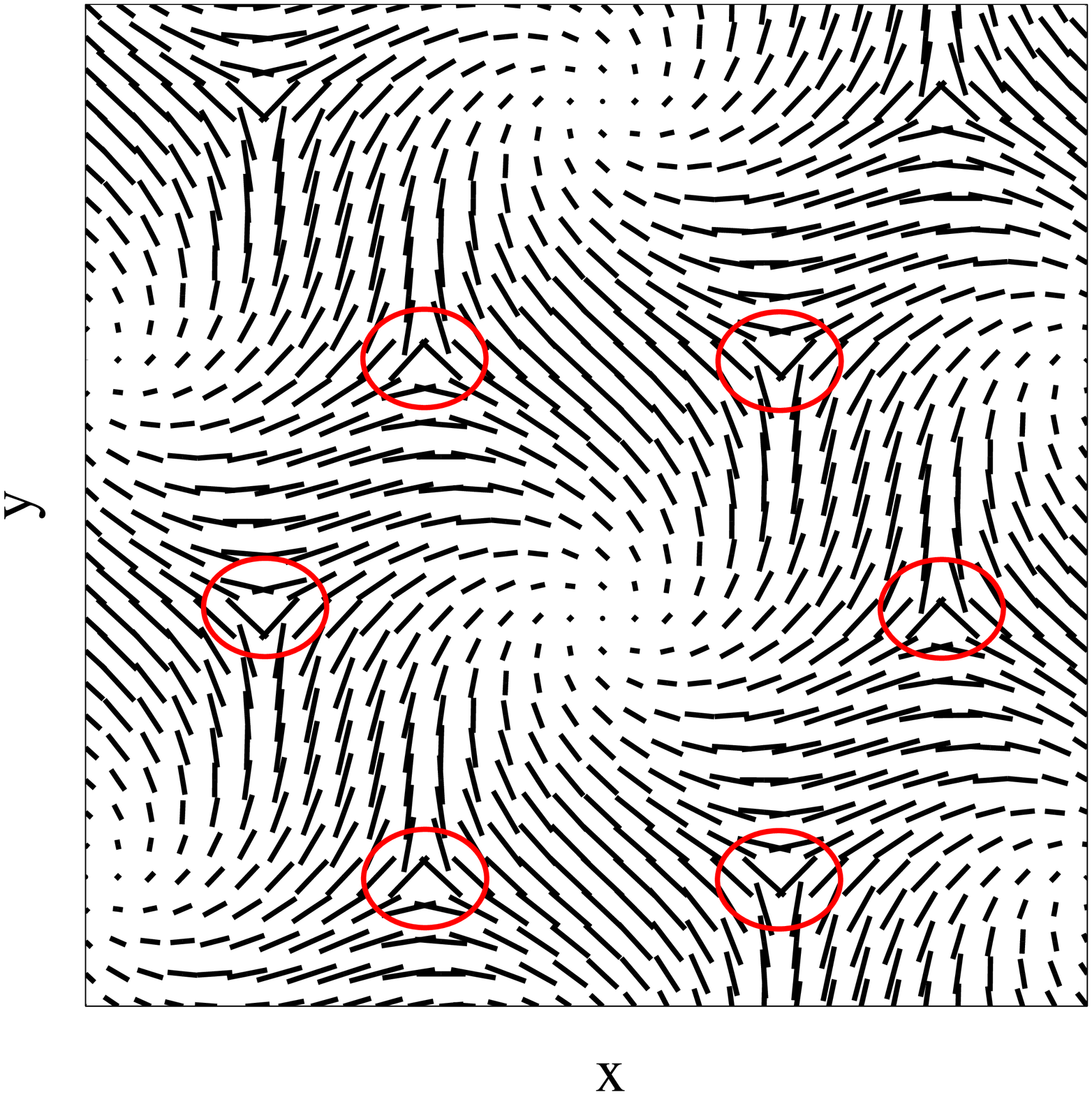}}
\caption{Director profile on the $xy$ plane taken at $L_z/2$ of the metastable state $f)$ of Fig.\ref{fig1}. Red circles indicate the presence of a charge $-1/2$ topological defect.} 
\label{fig4}
\end{figure*}

\begin{figure*}
\centerline{\includegraphics[width=18.cm]{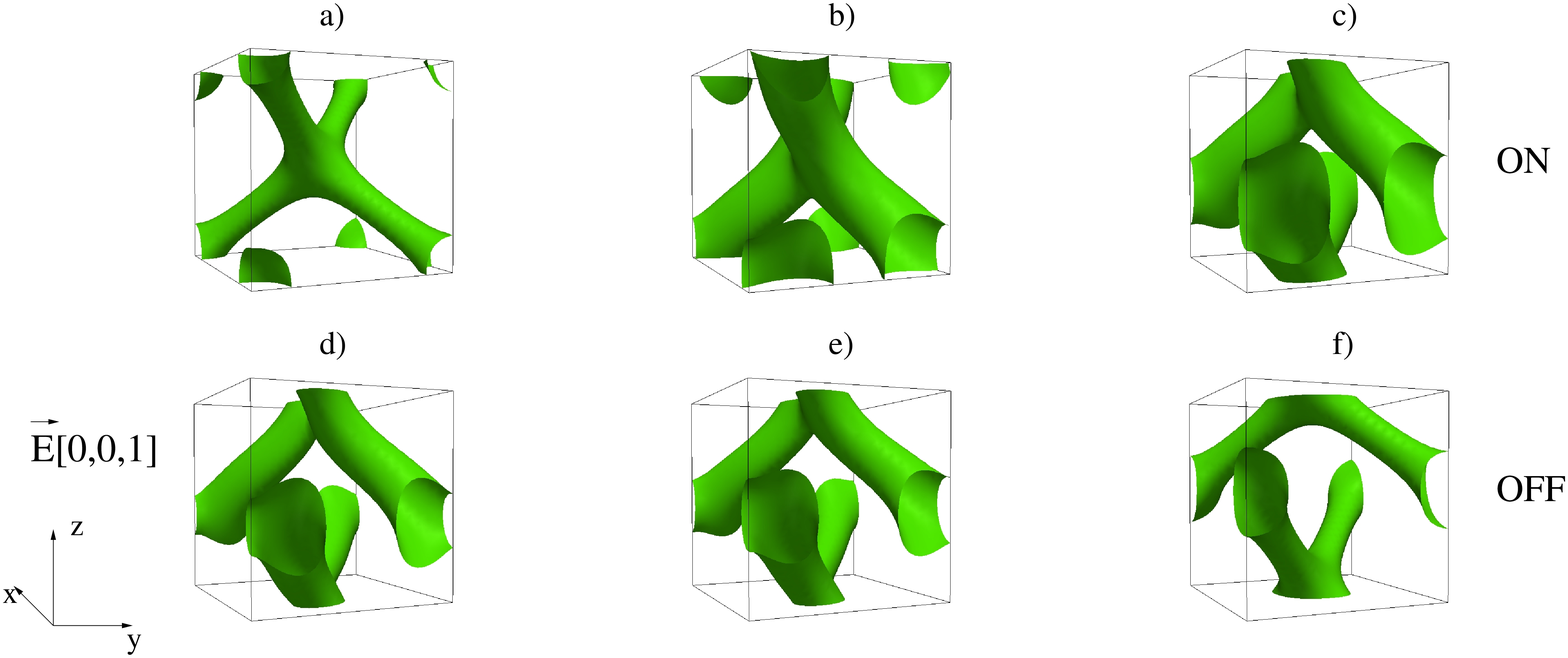}}
\caption{Switching defect dynamics of a BPII phase of unit cell size $32\times 32\times 32$ with periodic boundary conditions along all
directions. The electric field, applied along the $z$-direction ($[0,0,1]$), is ${\cal E}_{fl}^b \simeq 0.5$ 
(from $a)$ to $c)$). It is switched off at $t=80\times10^4$ (from $d)$ to $f)$). After one cycle, the system recovers the initial disclination structure 
(shifted along the $z$-direction), making the system switchable.}
\label{fig2}
\end{figure*}

As Fig. 1 clearly shows, the system cannot recover the initial state of 
Fig.\ref{fig1}$a$, but  it gets stuck into a metastable state (see Fig. 
\ref{fig1}$l$) whose free energy is higher than the initial one. 
On the other hand this design can be still considered as a switchable 
device since  further simulations show that there exists a reversible cycle   
between the field-induced state in Fig. 1$f$ and the metastable state in Fig. 1$l$.
It is interesting to look also at the  director profile of the state of  
Fig.\ref{fig1}$f$  induced by the flexoelectric coupling.  In
Fig.\ref{fig4} this is shown for a cross section  in the  $xy$ plane taken at $L_z/2$: the structure 
consists of an hexagonal lattice of defects of $-1/2$ topological charge that separate regions of splay-bend distortions around a line passing through the center of the hexagon (i.e., the disclination in three dimensions).
This spatial organization of the defects  is in agreement with what found  in Refs.~\cite{alex3,ef_theor2}, 
where a similar structure has been identified  as a stable  configuration of the director field, for 
either a spontaneously splayed nematic liquid crystal or a blue phase under an intense electric field,
respectively. We find here that this state is naturally found in blue phases with flexoelectricity, and
hence should be particularly relevant for the stabilised BPs recently found experimentally in
Ref.~\cite{Coles_Pivnenko:2005:Nature}.

A different kinetic pathway is observed when a field (still along the $z$-direction  with ${\cal E}_{fl}^b \simeq 0.5$) is applied on a BPII structure with periodic boundary conditions (see Fig.\ref{fig2}).  With respect to the BPI case 
the effect of the field is less dramatic and the typical 
(zero field)  BPII  network of defects  is still  recognizable even when the field is on. 
The persistence of a structure resembling (or topologically connected with) that of zero field, 
arguably suggests that the device should switch back to BPII when the field is taken off. This is indeed what happens.
More precisely,  when the field is switched on, the defect structure is initially bent and twisted at  its center, while the regions in which the director field is distorted around a defect line  widens (Fig.\ref{fig2}$b$). The latter effect is made apparent by looking at the effective thickness of the disclination tubes. The device is driven towards a field induced state characterised by a defect structure similar to the zero-field (equilibrium) one (the shift along the $z$-direction is immaterial in view of the periodic boundary conditions, see Fig.\ref{fig2}$c$).
When the field is switched off, the disclination profile remains essentially undistorted, with only a progressive slow reduction of the thickness of the defects (Fig.\ref{fig1}$d$ and 
$e$) until the typical zero-field BPII structure (shifted along the $z$-direction) is recovered. 
Notice that only one cycle is here necessary to restore the initial 
defect pattern. 

\begin{figure*}
\centerline{\includegraphics[width=17.5cm]{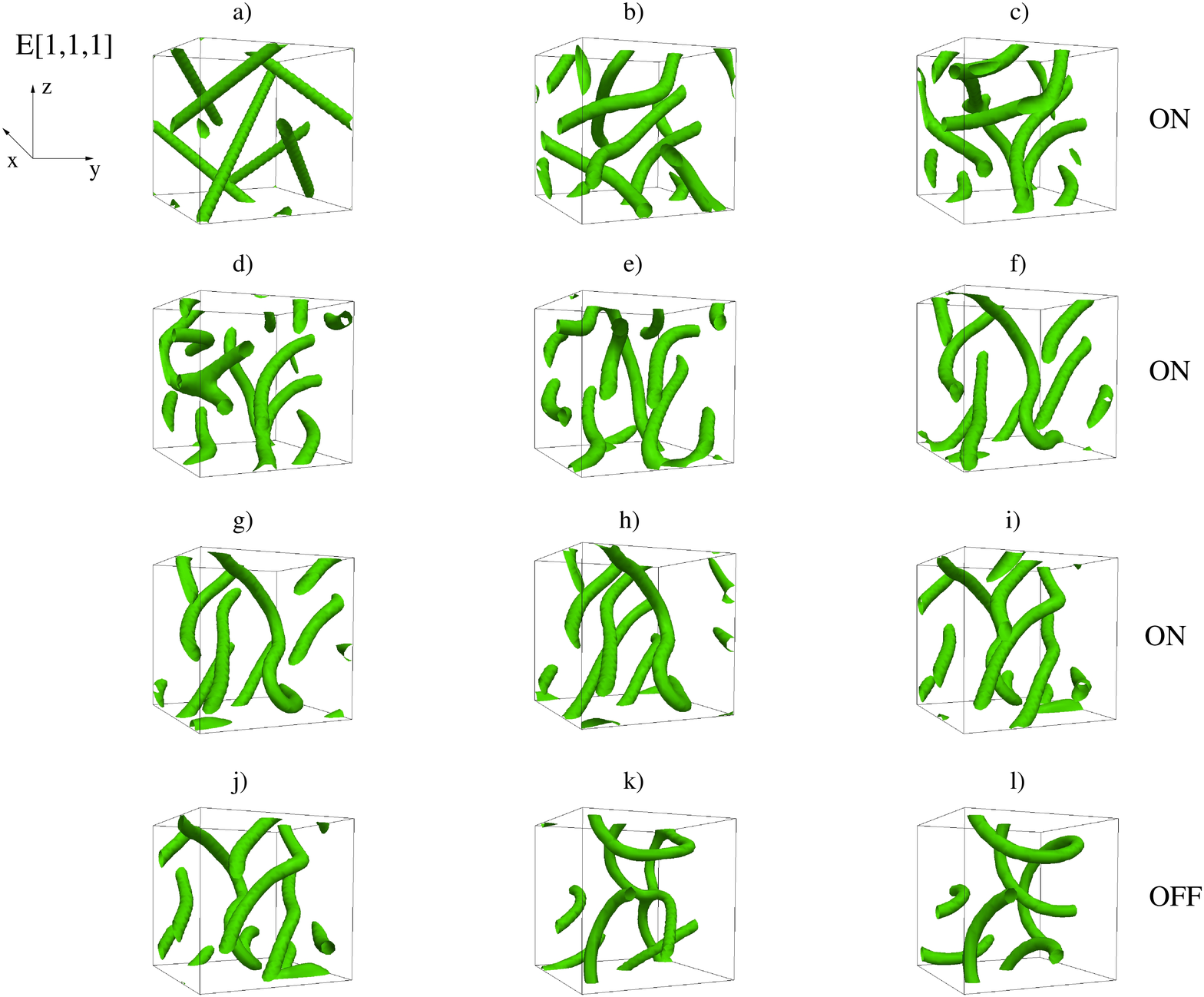}}
\caption{Evolution of the disclination network of the BPI phase during the switching on ($a)-c)$)
-off ($d)-f)$) dynamics in a simulation box of size $32\times 32\times 32$ with periodic boundary conditions in all directions. 
The electric field, applied along the diagonal direction ($[1,1,1]$), 
is ${\cal E}_{fl}^b \simeq  0.5$. It is switched off at $t=60\times 10^5$.
When the field is on, the device is driven towards a metastable state. When the field is removed again a metastable state feauturing twisted disclinations is attained. }
\label{fig3}
\end{figure*}

We now look at the dependence of the defect dynamics with respect to the direction of the applied field.  
In Fig.\ref{fig3}  we report the time evolution of the defect network of BPI when a field is applied along
the diagonal of the unit cell direction ($[1,1,1]$) which is not related to
$[0,0,1]$ by the cubic symmetry of the structure. 
Similarly to the case with the field along the $z$-direction, the strength of the applied field is large  enough to significantly rewire the defect network. 
The steady-state field-induced state still has the defect lines broadly aligned along the field, but not as markedly as in the $[0,0,1]$ case.
When the field is switched off, the device gets again stuck
in a metastable configuration, different than the one obtained when 
removing the field along the $[0,0,1]$ direction. 

\begin{figure*}
\centerline{\includegraphics[width=19.cm]{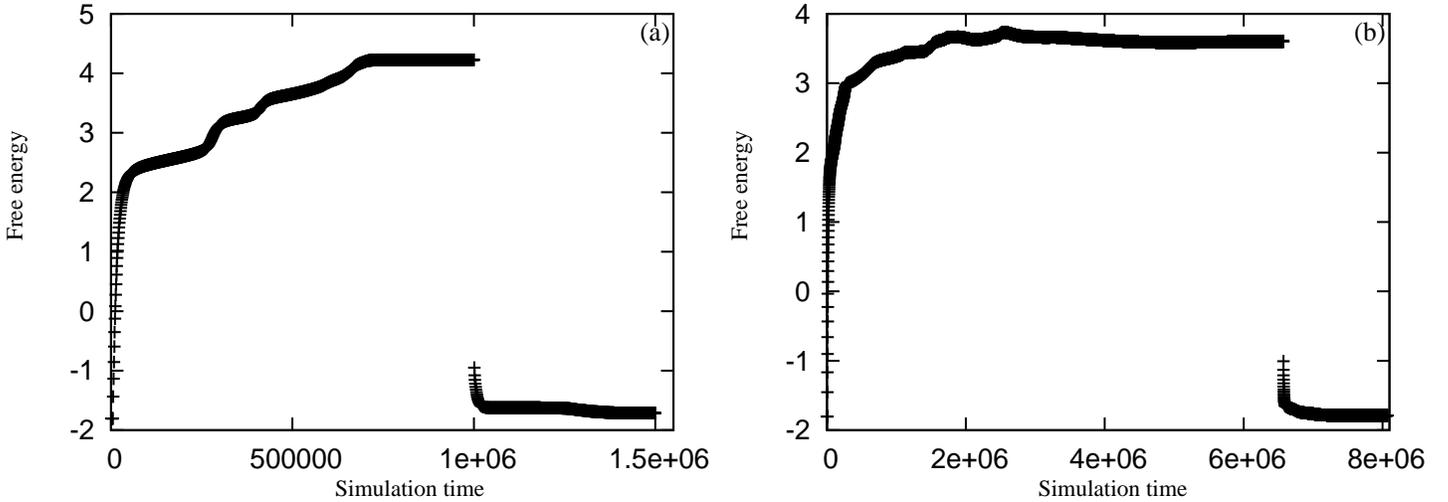}}
\caption{Time evolution of the free energy in a BPI unit cell
during a switching on-off dynamics with an electric field applied along the $z$-direction ($[0,0,1]$) (plot $(a)$) and along
the diagonal direction ($[1,1,1]$) (plot $(b)$). 
The field is ${\cal E}_{fl}^b \simeq 0.5$ in both cases.
Interestingly the time evolution
of the diagonal field case is $\sim 6$ time longer than the case in which the field is along the $z$-direction. (These free energy curves correspond to
runs without hydrodynamics. This omission allows us to reach a larger
value of the electric field, which emphasises the differences in the
dynamics, avoiding at the same time numerical stability issues).}
\label{fig5}
\end{figure*}

Intriguingly, we find the switching-on dynamics depends strongly on field 
direction, and in all cases proceeds via multiple steps
visible in the transient plateaus observed in the plots of the
free energy versus time (see Fig.~\ref{fig5}).

\subsection{BP samples under confinement}

We now turn to the switching dynamics of a  field in confined BP samples. 
These are particularly relevant to real-world devices, as these necessarily include boundaries, where the director field can easily be anchored, either chemically or mechanically via rubbing -- and the anchoring strength can often be controlled.
In this Section we will only focus on BPI cells, which, as shown in the previous Section, lead to a more interesting dynamics in the bulk. 
We consider either homeotropic (normal) or homogeneous (planar) anchoring on upper and lower walls, and periodic boundary conditions in other directions. 
To better compare with the results discussed previously, the size of the simulation box is set to $32\times 32\times 32$. The 
field is then applied along the $z$-direction ([0,0,1]). Once more, its magnitude is strong enough that the initial defect network breaks up.
The case of a surface flexoelectric coupling is also explored below -- in this case the field effects are confined to the boundaries.

The presence of confining surfaces strongly affects the stability of defects in liquid-crystalline systems. 
In previous numerical studies~\cite{fukuda1,fukuda3}, the existence of several defect structures (sometimes described as \lq\lq exotic\rq\rq structures), very different from that of the equilibrium BPI cell, have been proposed. 
A parallel array of double-helical disclinations, and two sets of undulating defect lines, were observed in a cell with homeotropic anchoring on both (parallel and flat) walls, when their distance was of the order of the helical pitch~\cite{fukuda1}. On the other hand, ring defects structures emerge when
homogeneous anchoring is set on both walls~\cite{fukuda3}. A key question, crucial for devices, concerns how to dynamically switch among different metastable states to kinetically reach, in a controlled way, a target structure. 
Recent studies~\cite{atsoft} on the switching defect dynamics 
showed that applied electric fields could represent an interesting route to reveal the existence of even more metastable states.
A simple switching on-off schedule can lead, for example, to a bistable defect network~\cite{prl}, where each state is uniquely selected by the application of the electric field along an appropriate direction.
Interestingly, some key questions arise when flexoelectric in confined samples is taken into account. What is the switching dynamics? Is this affected by the choice of different boundary conditions? 

\begin{figure*}
\centerline{\includegraphics[width=18.cm]{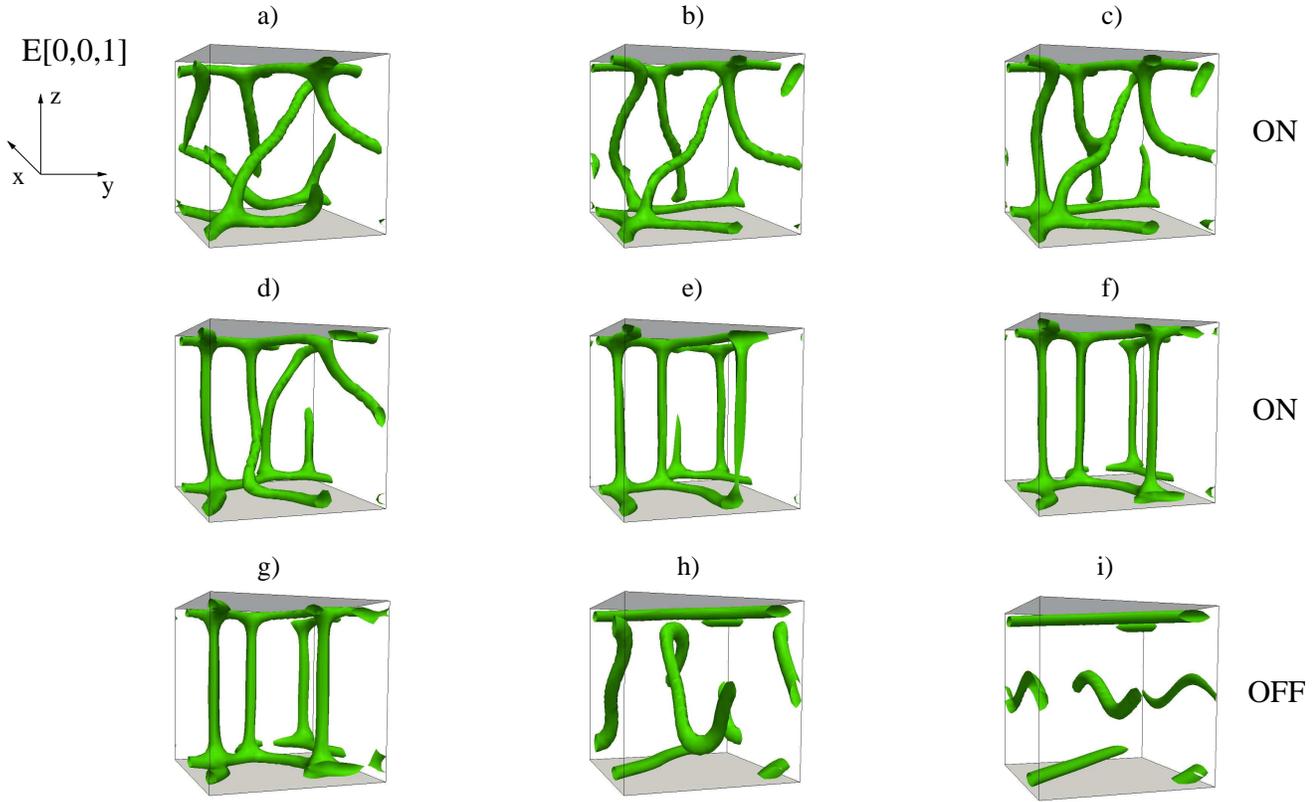}}
\caption{Evolution of the BPI disclination network in a confined sample of size $32\times 32\times 32$ with homeotropic anchoring. Top and bottom walls (in grey) are oriented in the $xy$ plane.
The switching on is from $a)$ to $f)$ and the switching off is from $g)$ to $i)$.
The field, applied along the $z$-direction ($[0,0,1]$), is such that ${\cal E}_{fl}^b \simeq 0.5$. It is
switched off at $t=100\times 10^4$. The switching on drives the device towards a field induced state $f)$ in which column-like defects are connected at the walls,
while, when the field is removed, the cell gets stuck into a metastable state with straight disclinations close to the walls coexisting with undulating lines in the middle of the BPI cell.}
\label{fig6}
\end{figure*}

To address these issues, in Fig.\ref{fig6} we show the flexoelectric switching of a BPI cell with homeotropic anchoring (with a slight pretilt of  $\sim 10^{\circ}$ from the $z$-direction) on both walls.
We have also set ${\cal E}_{fl}^b \simeq 1$ (applied along the $z$-direction) and $W_0=0.1$, which still corresponds to moderately strong anchoring.  

The first thing to note is that the equilibrium defect structure (Fig.\ref{fig6}$a$) is different from the typical one observed for the BPI at equilibrium 
with periodic boundaries, although the basic BPI topology is still recognisable 
(for instance in the disclinations in the interior of the cell).
This metastable structure is the same as that reported in Refs.~\cite{fukuda1,prl}. A second, key, observation is that the dynamics of flexoelectric switching are significantly affected by the anchoring. Thus, the normal anchoring
conditions in Fig.~\ref{fig6} lead to a visible initial bend 
of the diagonal disclination. This is followed later on by the formation
of straight defect lines as in the bulk case (see Fig.\ref{fig6}$b$ and $c$),
which branch out at the boundary to end up parallel to the confining planes
there (see Fig.~\ref{fig6}$d$--$f$). 
Once more, upon field removal the system drifts away from the field-induced
state in Fig.~\ref{fig6}f into another metastable state, featuring undulating
defect lines close to but distinct from those observed in the bulk (compare
Fig.~\ref{fig6}i with Fig.~\ref{fig1}l).

\begin{figure*}
\centerline{\includegraphics[width=18.cm]{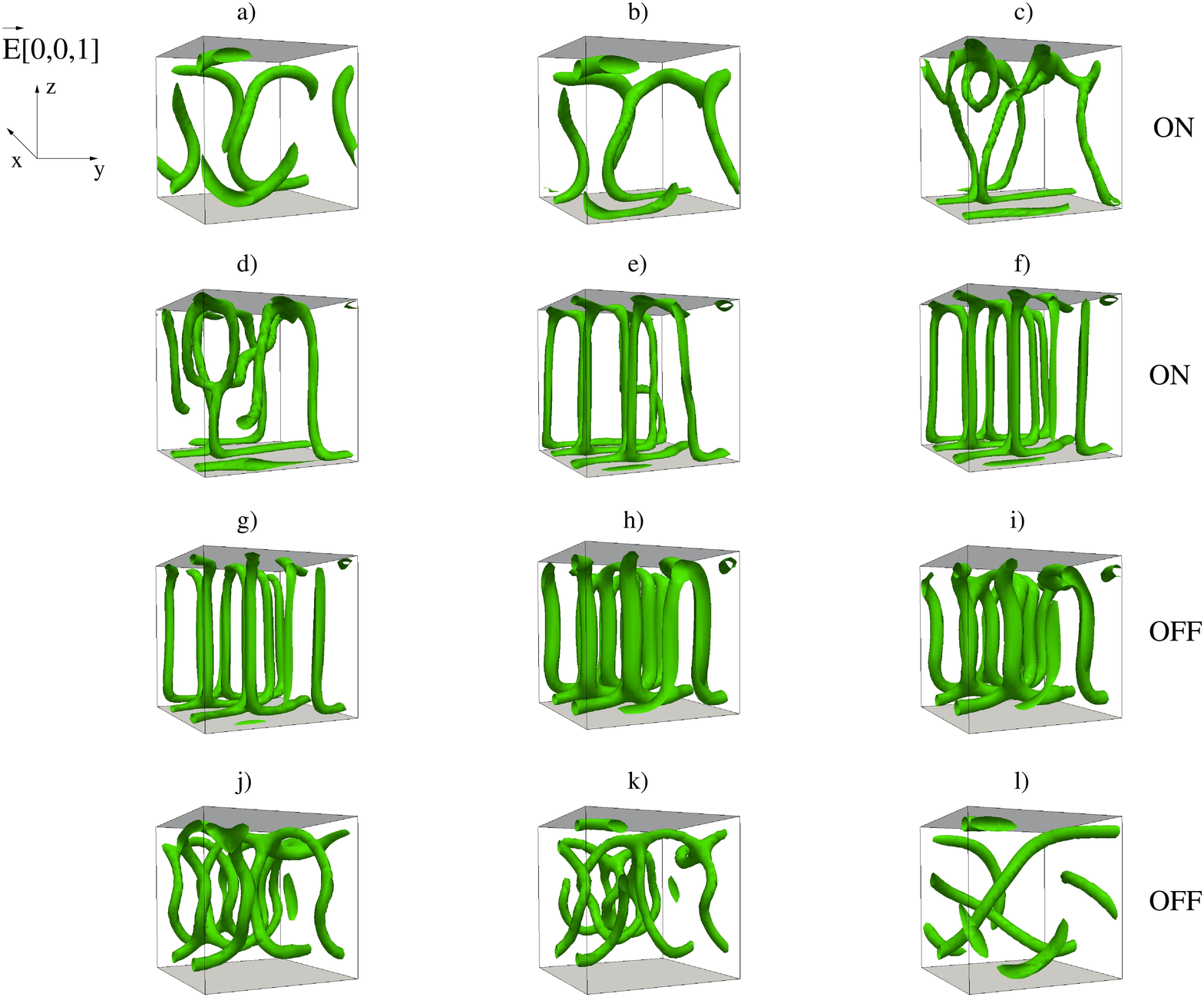}}
\caption{Evolution of the BPI disclination network in a confined sample of size $32\times 32\times 32$ with homogeneous anchoring. Top and bottom walls (in grey) are oriented in the $xy$ plane. The switching on is from $a)$ to $f)$ and the switching off is from $g)$ to $l)$. The field, applied along the $z$-direction ($[0,0,1]$), is such that ${\cal E}_{fl}^b \simeq 1$. It is switched off at $t=128\times 10^4$. Twisted ring defects are recognisable during the switching on, while a novel metastable state $l)$, with arcs of defects spanning the device, is obtained when the field is removed.}
\label{fig7}
\end{figure*}

Homogeneous boundary conditions, as can be achieved by rubbing the liquid crystal at the boundaries, lead to yet different structures in equilibrium, and to new kinetic pathways during switching on and off a field. The behaviour of a BPI cell with such planar anchoring (along the $x$ direction) is shown in Fig.~\ref{fig7}. The flexoelectric coupling is again given by ${\cal E}_{fl}^b \simeq 1$, and the anchoring $W_0=0.1$ as in Fig.~\ref{fig6}.
The equilibrium defect network is now made up of an array of isolated twisted ring disclinations located in the bulk of the device (Fig.\ref{fig7}$a$). These structures have already been discussed in Ref.~\cite{fukuda3} and, as a field induced state, in the bistable BP device proposed in Ref.~\cite{prl}: we do not dwell on them here, and rather focus on the field-induced aspects. 
When the field is switched on, the network is initially bent and stretched (Fig.\ref{fig7}$b$ and $c$) with disclinations again alinging parallel to the field until they touch the walls. 
Later on they recombine into elongated rings now connected to defects pinned up at the walls 
((Fig.\ref{fig7}$d$, $e$ and $f$).
As in the previous cases, when the field is removed defects depin from the surfaces and form connected ring arcs in the bulk 
(from Fig.\ref{fig7}$g$ to $j$). Then the network develops branching points which are unstable and annihilate in the bulk leaving a state with arcs of defects spanning the whole device (Fig.\ref{fig7}$k$ and $l$).

Lastly, we investigate the switching dynamics in response to surface flexoelectricity. For strong anchoring these contributions can be rewritten as surface terms, so one may expect that their presence may only
lead to deviations from the equilibrium bulk disclinations 
in strongly confined samples, such as the ones we are considering. 
As we mentioned previously, because the field couples with the derivative of the order parameter (see Eq.\ref{surf_freeen}), in the case of strong anchoring the coupling is identically zero at the boundaries and the surface flexolectric 
term is not effective.
However for finite anchoring (as the case of $W_0=0.1$ we consider), this is not true anymore, and the director can deviate from the anchoring direction, at the cost of some surface energy.

\begin{figure*}
\centerline{\includegraphics[width=18.cm]{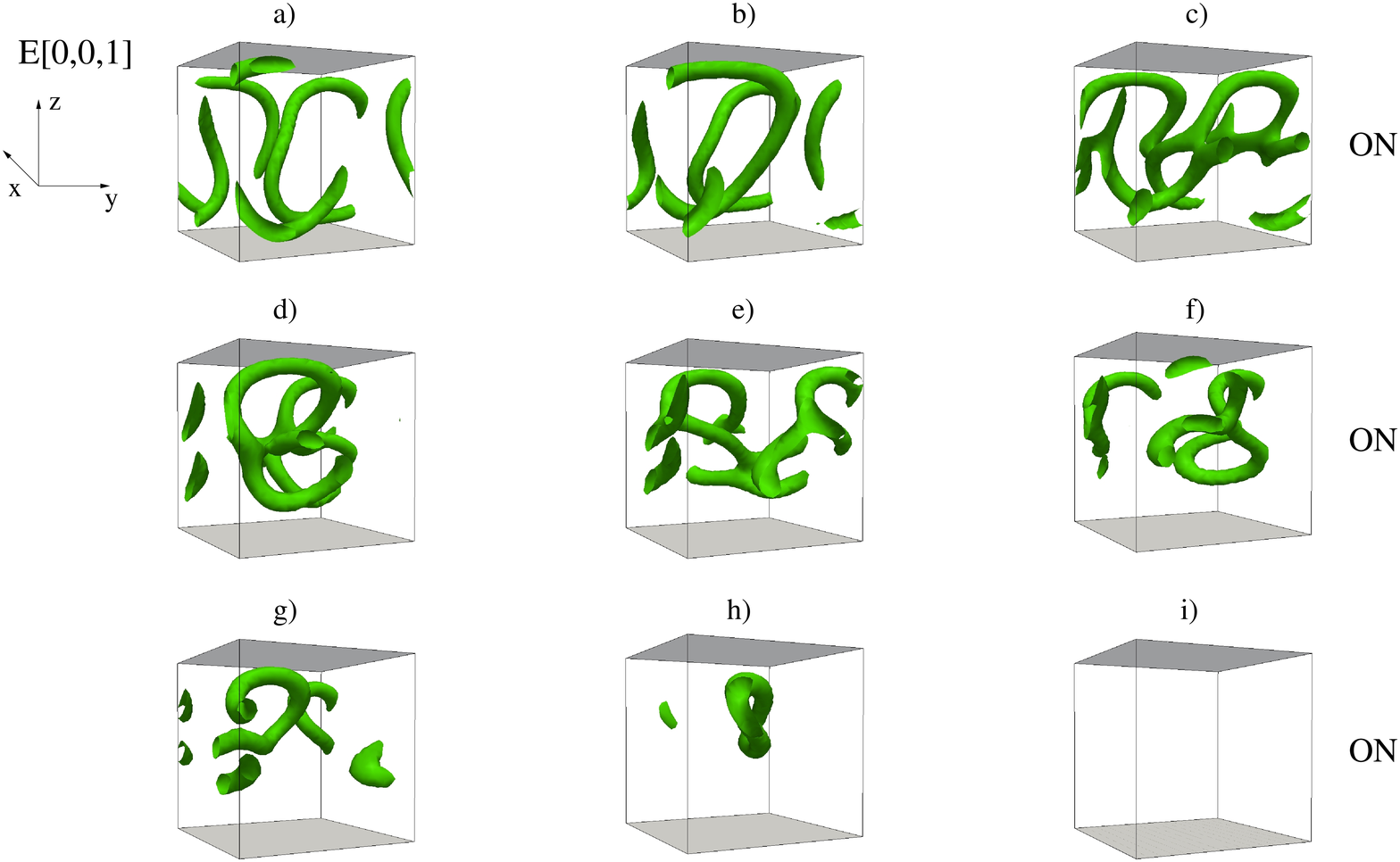}}
\caption{Evolution of the BPI disclination network in a confined sample of size $32\times 32\times 32$ with homogeneous anchoring and with surface flexoelectric coupling. 
Top and bottom walls (in grey) are along $xy$ plane. The switching on is from $a)$ to $i)$, where all defects disappear. 
The field, applied along the $z$-direction ($[0,0,1]$), is such that ${\cal E}_{fl}^s \simeq 0.15$.}
\label{fig8}
\end{figure*}

\begin{figure*}
\centerline{\includegraphics[width=18.cm]{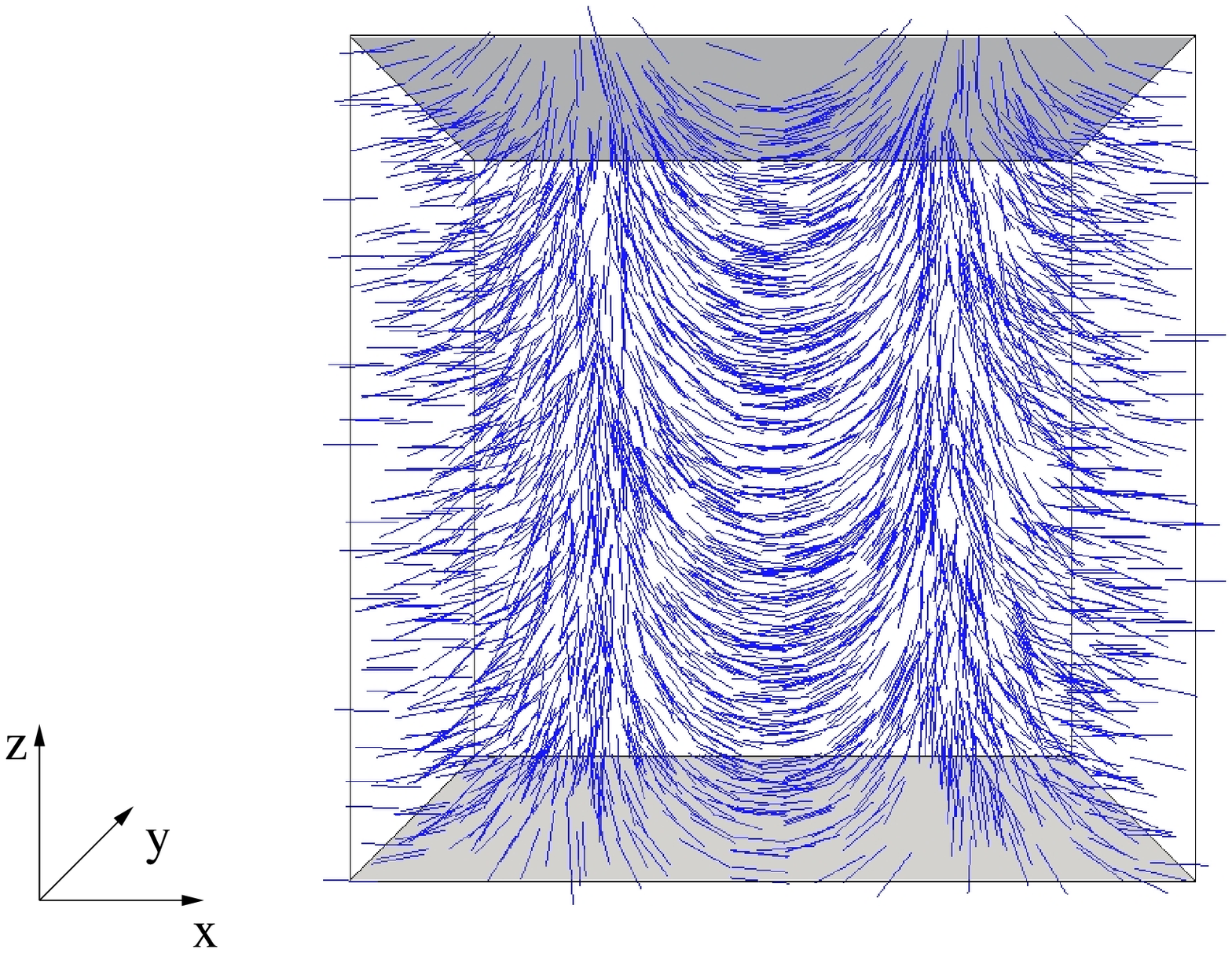}}
\caption{3D director profile of the field-induced defect-free state in Fig.~\ref{fig8}i.}
\label{fig9}
\end{figure*}

To address the role of the surface flexoelectric coupling, we  simulated the evolution of disclinations in a confined BPI cell with homogeneous anchoring (Fig.\ref{fig8}). 
The magnitude of the field, applied along the $z$-direction, is such that ${\cal E}_{fl}^s \simeq 0.15$ (a pretilt of $\sim 10^{\circ}$ from the plane is also included). Now the switching leads to a shrinkage and eventual annihilitation of the twisted disclination loops (the full kinetic pathway is shown in Fig.\ref{fig8}a-i). The final state is defect-free (Fig.~\ref{fig9}), and the associated director profile has a lot of splay-bend deformations. Intriguingly, this state is similar to the spontaneously splayed nematics in an achiral liquid crystal with a strong flexoelectric coupling, predicted in~\cite{alex3}. 
As expected, the state remains disclination free even when the field is removed.
 
\section{Conclusions}

To summarise, we have numerically investigated the switching dynamics of cholesteric blue phases in the presence of a flexoelectric coupling with an electric field.
The choice of boundary conditions (periodic boundaries, or confined samples with homogeneous or homeotropic anchoring on both walls) and of the dynamical schedule of the field application crucially affect the dynamics, yielding several intriguing defect structures. Some of these resemble the zero-field equilibrium network (e.g. for BPII), while others organize forming novel field induced stable or metastable states. The most frequently observed structure is one where the disclinations are aligned (more or less markedly, according to the different cases) along the field direction. 

With periodic boundary conditions (corresponding to the bulk of a large sample) 
and with a field applied along the $[0,0,1]$ direction, BPI defects
align along the direction of the applied field via
a complex kinetic pathway which includes the formation
of intermediate \lq\lq X-like\rq\rq disclination junctions. 
Once the field is switched off, the network is stuck in a zero-field metastable state.
Remarkably, a further application of the field restores the field induced states, thereby making these cells switchable.
When the field is applied along the $[1,1,1]$ direction, the alignment is
imperfect and the dynamics is slower.

Importantly, in both cases, as for most of the other geometry/boundary condition choices, upon switching off, the field-induced BPI structure drifts away from the field-induced conformation and gets stuck into a metastable phase, the details of which depend on boundary conditions, field directions etc.. This means that, while the field-aligned disclination state is switchable, it cannot, at least under the conditions explored by our simulations, be used as the basis for multistable energy-saving devices such as those described in~\cite{prl}. (However we have laid the formulation for future 
computational searches for these). 
The inability to reform an equilibrium BPI structure also suggests that large memory effects are present.

Confined samples lead to qualitatively different results. With normal, or homeotropic, anchoring at the boundaries, and bulk flexoelectric coupling only, the field-aligned disclinations branch at or close to the wall to end up parallel to the boundaries there.
With homogeneous anchoring the defects without a field form twisted rings; these elongate and align with the field when bulk flexoelectricity is switched on. 
Surface flexoelectric contributions, relevant at low to intermediate anchoring strength, can drive the device into yet more configurations, such as a strongly splayed-bent nematic conformation (see Fig.~\ref{fig9}).

Our results give, we believe, a convincing demonstration that flexoelectricity and geometry
are important factors in selecting defect networks in blue phases. Given the
importance of flexoelectric coefficients in the materials used to
build the new blue phases used in Ref.\cite{Coles_Pivnenko:2005:Nature}, 
we hope our results will be useful when designing new blue phase-based
liquid crystal devices.

\vskip 1truecm

\section*{Acknowledgements}

We thank EPSRC EP/J007404 for funding. MEC holds a Royal Society Research Professorship.

\end{document}